\documentclass[prb,twocolumn,preprintnumbers,amsmath,amssymb,longbibliography]{revtex4}

\usepackage{graphicx}
\usepackage{color}

\begin{document}

\title{Deformation and scattering in graphene over substrate steps}
\author{T. Low, V. Perebeinos, J. Tersoff and Ph.\,\,Avouris}
\affiliation{IBM T.J. Watson Research Center, Yorktown Heights, NY 10598, USA}
\date{\today}

\begin{abstract}
{
The electrical properties of graphene depend sensitively
on the substrate.  For example, recent measurements of
epitaxial graphene on SiC show resistance arising from
steps on the substrate.
Here we calculate the deformation of graphene at substrate steps,
and the resulting electrical resistance, over a wide range of step heights.
The elastic deformations contribute only a very small resistance at the step.
However, for graphene on SiC(0001) there is strong substrate-induced doping,
and this is substantially reduced on the lower side of the step
where graphene pulls away from the substrate.
The resulting resistance explains the experimental measurements.
\\
}
\end{abstract}

\maketitle

The advance of very high speed graphene
electronics\cite{avouris10,lin11} depends on understanding and controlling
the interaction of graphene with the supporting substrate.
Electron mobility can vary
over many orders of magnitude depending on
the substrate \cite{neto09,dean10,berger06}.
Among other factors, morphological deformations of the graphene
may limit mobility
\cite{dimitrakopoulos11,farmer11,bryan11,shuaihua,kim11}.
It is therefore important to determine how the substrate morphology
affects transport in supported graphene.

Epitaxial graphene on SiC provides an ideal system in which
to study the role of substrate morphology.
SiC is a promising substrate because,
in contrast to other approaches, it allows growth of
epitaxial graphene directly on an insulating substrate \cite{emtsev09}.
However, epitaxial graphene on SiC substrates generally exhibits
smaller carrier mobilities than exfoliated graphene
on SiO$_2$ substrates \cite{berger06,emtsev09,tedesco09,RWTC09}.
The reason for this difference is not fully understood,
but SiC substrates exhibit a high density of multilayer steps,
which are implicated in the lower mobility.
Several experiments show that resistance increases
with step density\cite{dimitrakopoulos11},
step heights\cite{shuaihua} and step bunching\cite{farmer11,bryan11};
and the local electrical resistance associated with individual substrate steps
has recently been measured \cite{shuaihua}, 
by scanning potentiometry in a
scanning tunneling microscope.

\begin{figure}[t]
\centering
\scalebox{0.86}[0.86]{\includegraphics*[viewport=250 293 500 450]{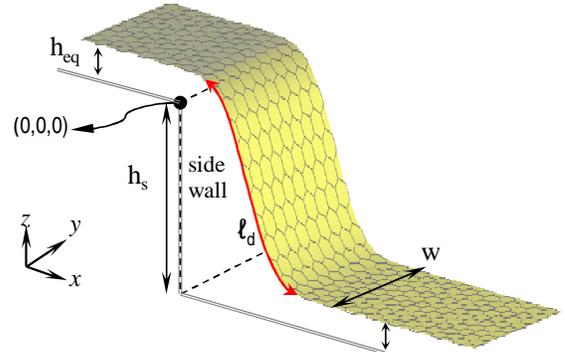}}
\caption{Illustration of graphene over a substrate step.
Here $h_s$ is the step height and
$h_{eq}$ is the equilibrium distance of graphene from the substrate surface
due to van der Waals interaction. $\ell_{d}$ is the length of graphene
detachment from the substrate.
}
\label{fig1}
\end{figure}

Here we study graphene over an abrupt substrate step,
as illustrated in Fig.\,\ref{fig1},
calculating both the structural deformation
and the resulting electrical resistance.
The results are directly applicable to epitaxial graphene on SiC,
and also show more generally
how the morphology affects electrical transport.
We find that very little resistance arises directly from the
structural deformations, despite the strong curvature of graphene
as it passes over a step.
For SiC, we nevertheless find a substantial resistance
associated with the step, in good agreement with experiment\cite{shuaihua}.
This resistance arises almost entirely from the
electrical coupling between the graphene and substrate,
which varies sharply in the vicinity of the step.
Thus morphology plays a qualitatively different
and far more important role in substrates such as SiC
that dope the substrate or otherwise couple strongly,
than it does for substrates such as SiO$_2$
that are electrically passive.

We begin by determining the graphene geometry as it passes over a substrate step.
The graphene deformation is determined by a balance between the van der Waals
interaction, which favors conforming to the substrate,
and elasticity, which favors keeping the graphene flat and smooth.
Since the displacement field can vary on the atomic scale,
we use an atomistic valence force model (VFM) to describe the
elastic deformations \cite{PT09}.
The van der Waals interaction between graphene and substrate is modeled
with the Lennard-Jones (LJ) 6-12 potential \cite{GL56}.
The parameters for our LJ model are determined by setting the
equilibrium distance between graphene and substrate to
$h_{eq}\approx 3.4\,$\AA \cite{mattausch07}, and the
binding energy to $E_{B}\approx 40\,$meV per atom \cite{GL56,ZUH04}.
The total energy is then simply the sum of these two contributions,
${\cal E}={\cal E}_{elas}+{\cal E}_{vdw}$.

\begin{figure}[t]
\centering
\scalebox{0.86}[0.86]{\includegraphics*[viewport=217 360 530 550]{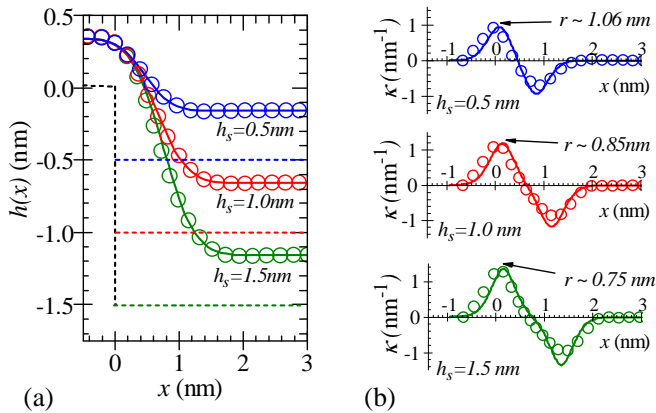}}
\caption{\footnotesize  $\bold{(a)}$
Graphene geometry $h(x)$ for various step heights as indicated.
Symbols are the full numerical calculation.
Solid lines are best fit using Eq.\,\ref{errfunch}.
The respective step profile are illustrated by dashed lines.
$\bold{(b)}$ Corresponding curvature $\kappa$ for
geometries shown in (a), with lines and symbols as in (a).
The minimum radius of curvature (inversion of maximum $\kappa$) is indicated.
}
\label{fig2}
\end{figure}

We calculate the minimum-energy geometry
for graphene over a wide range of step height $h_{s}$,
allowing the graphene to slide to relax any in-plane strain.
The relaxed geometries $h(x)$ for three step heights typical of SiC
are plotted in Fig.\,\ref{fig2}a. \textcolor{black}{The presence of the atomic step increases the graphene area compared to
when the substrate is flat. The extra length increases
with $h_{s}$ and was found to be $0.1$, $0.3$, and $0.6\, $nm respectively.
The nonlinear dependence reflects the increasing distortion (steeper slope) with increasing step height.}
The maximum slope is of order 1, confirming the need for
a fully numerical treatment.
% $1$, $3.2$, $6.2\,\AA$

We find that the calculated geometries
can be well approximated by
a simple error function,
\begin{eqnarray}
h(x)\approx-\frac{h_s}{2}\left[\mbox{erf}\left(\frac{x-x_s}{d_s}\right)+1\right]+h_{eq}
\label{errfunch}
\end{eqnarray}
As shown in Fig.\,\ref{fig2}b, even the variation in curvature
across the step is well described by this simple functional form.
The only noticeable discrepancy is that the error function
is symmetric, while in the full calculation there is a slight asymmetry,
with the maximum curvature induced at the upper edge of the step. For step height $h_{s}=1.5\,$nm, we find a maximum curvature
equivalent to that of a carbon nanotube of diameter $1.5\,$nm.

\begin{figure}[t]
\centering
\scalebox{0.76}[0.76]{\includegraphics*[viewport=185 255 545 445]{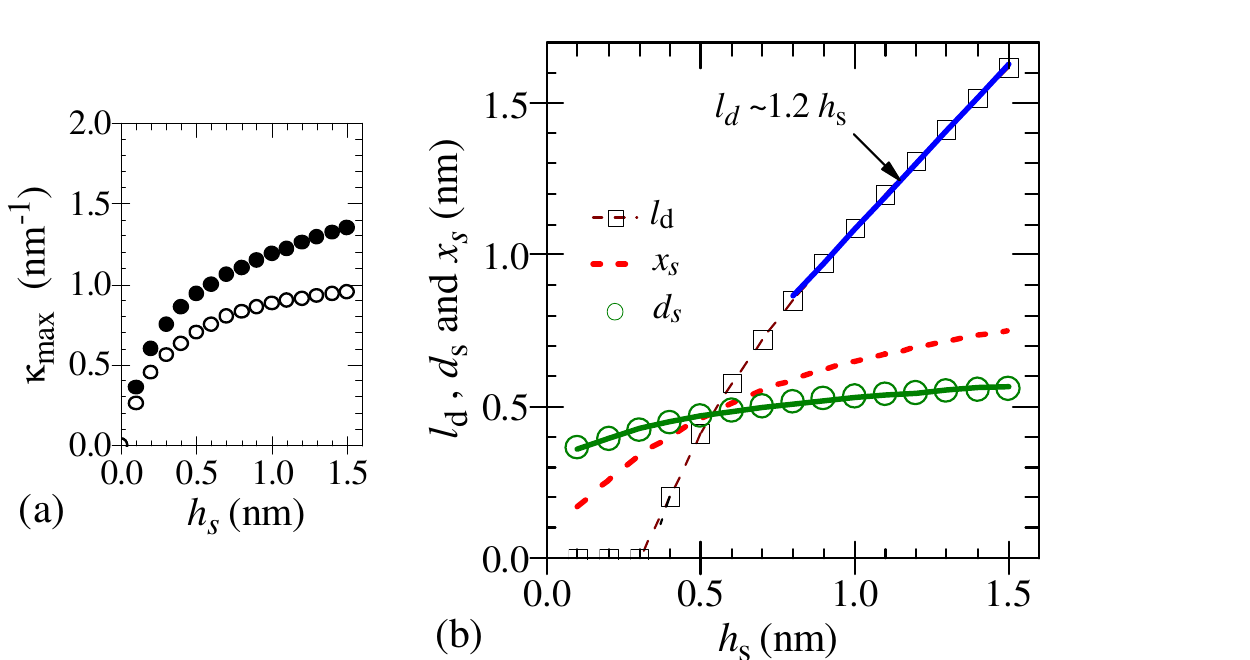}}
\caption{\footnotesize 
$\bold{(a)}$ Numerically calculated maximum curvature
$\kappa_{max}$ for the upper (solid) and lower (open) edge of the step
as function of step height $h_s$. 
Note that the often used approximation for curvature,
$\kappa\approx\partial_{x}^{2}h$, is only reasonable for $h_s$ of few angstrom.
Over the range of $h_s$ of our interests, it grossly overestimates
the real curvature due to the large gradients in our graphene geometries.
$\bold{(b)}$
Dependence of graphene deformation on step height $h_s$.
Graphene step width $d_s$ and lateral shift $x_{s}$ are
obtained by fitting Eq.\,\ref{errfunch} to the relaxed geometry
$h(x)$ obtained numerically.
We also show the detachment length $\ell_{d}$, i.e.\ the length of graphene
separated from the substrate by $2h_{eq}$.
Note that $\ell_{d}\approx 1.2 h_{s}$ for large $h_s$.
The step width $d_{s}$ depends only weakly on step height.
}
\label{fig3}
\end{figure}

The maximum curvature $\kappa_{max}$ as a function
of step height $h_s$ is plotted in Fig. \ref{fig3}a,
for both the lower and upper edge of the step.
It is proportional to $h_s$ in the small $h_s$ limit.
For the lower edge, the limit of large $h_s$ corresponds
to the well-known problem of peeling \cite{landau86}.
In this limit, $\kappa_{max}$ approaches
$\sqrt{2E_{B}/a\beta}\approx 1.2\,$nm$^{-1}$, where
$a$ is area per carbon atom and $\beta\approx 2.1\,$eV \cite{PT09} is the bending
rigidity.

Figure \ref{fig3}b summarizes the dependence of
the graphene deformation $h(x)$ on step height $h_{s}$,
in terms of the parameters of Eq.\,\ref{errfunch}.
For a given step height, 
the characteristic step width $d_s$
determines the maximum radius of curvature $r=1/\kappa_{max}$.
Another relevant lengthscale is the length $\ell_{d}$
over which the graphene is detached from the substrate.
For concreteness we define $\ell_{d}$ to be the length of graphene
separated by $2h_{eq}$ or more from the substrate surface.
Fig.\,\ref{fig3}b shows that $\ell_{d}$ remains zero at small
$h_s$ but begins to increase rapidly for $h_s > h_{eq}$.
At larger step height we find that $\ell_d \sim 1.2\, h_{s}$,
which proves important in later discussions.

\emph{Geometry-induced scattering:}
Curvature results in bandstructure changes that can scatter
electrons near the step. \textcolor{black}{ To examine
this effect, after performing the geometry relaxation, we construct the Hamiltonian ${\cal H}$
within a nearest-neighbor Slater-Koster
parameterized $sp^{3}$ tight-binding model \cite{TL88}.  We then calculate the transmission and the electrical resistance ${\cal R}$
using the non-equilibrium Green's function formalism\cite{D97,DV08} in the limit of small voltage across the step and no inelastic scattering at the step.  We use the
known Fermi level\cite{bostwick07,mattausch07} of $E_{f}=0.45\,$eV
for graphene on SiC (0001).}

\begin{figure}[t]
\centering
\scalebox{0.9}[0.9]{\includegraphics*[viewport=260 160 505 445]{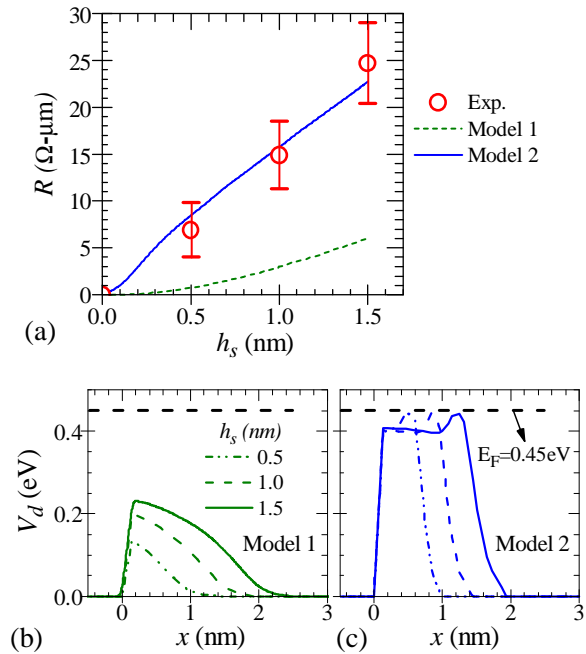}}
\caption{\footnotesize $\bold{(a)}$ Resistance through graphene due to substrate step, ${\cal R}$,
as function of step height $h_s$.
Various electrostatic models considering doping variations are plotted,
as described by Eq.\,\ref{chargecon} (Model 1) and
Eq.\,\ref{dft} (Model 2).
Experiment data\cite{shuaihua} is plotted in symbols,
with statistical error bar indicated.
$\bold{(b)}$ Potential profile $V_{d}(x)$ derived from Eq.\,\ref{chargecon} (Model 1)
for $h_0=0.5$, $1.0$, $1.5\,$nm. Red dashed line indicates the
Fermi level position.
$\bold{(c)}$ Same as (b), but derived from Eq.\,\ref{dft} (Model 2).
The small hump observed for $V_{d}(x)$ near the top is an artifact of
the polynomial form of $\Delta_{c}(h)$.
}
\label{fig4}
\end{figure}

Since the maximum curvature increases with step height $h_{s}$,
the resistance also increases.
For a step height of $1.5\,$nm, we obtain a resistance
$\sim 0.01\,$$\Omega$-$\mu$m.
Figure\,\ref{fig4}a includes results of a recent experiment\cite{shuaihua},
which employed scanning potentiometry in a scanning tunneling microscope
to resolve the resistance in graphene due to individual substrate steps.
The measured resistance ${\cal R}_{exp}$ has roughly
linear dependence with step height $h_s$,
$\sim 10\, \Omega$-$\mu$m for each nanometer step height.
Evidently, the resistance due to
curvature alone cannot account for this large ${\cal R}_{exp}$.
While $\pi$-$\sigma$ hybridization can result in
new scattering states in the vicinity of Dirac point \cite{blase94},
this effect is significant only when $r\lesssim 5\,$\AA \cite{blase94}.
Even for a very high step of $h_{s}=1.5\,$nm,
we find that the minimum radius of curvature
only shrinks to $r\approx 5\,$\AA\, if we assume
a much stronger van der Waals attraction,
with a binding energy $80\,$meV, which seems unphysical.

In our calculations thus far, we have ignored
possible in-plane strain inhomogeneities, which
is known to result in electron scattering \cite{fogler08,pereira09}.
Due to the different thermal expansion coefficient
between graphene and SiC, graphene can acquires a
residual biaxial strain upon cooling if sliding is suppressed.
Then due to the nonplanar geometry, graphene at the step
could experience a uniaxial stress relative to the rest of the sheet.
Graphene on SiC substrates is reported to have strains from $0.1-1\%$ as
measured by Raman spectroscopy \cite{RPSS09}.
To estimate a very conservative upper bound for ${\cal R}$ due to
strain inhomogeneity, we consider a local tensile strain of $1\%$ along
the detached region, with the step along the zigzag direction
where the scattering effect is largest \cite{fogler08}.
The result is ${\cal R} < 1\,\Omega$-$\mu$m.
Thus some source of scattering much stronger than the
geometrical deformations must be present
to account for the measured resistance.

\emph{Electrostatic doping effects:}
It is well known that when contacting graphene with  metals,
a difference in workfunction results in
electrostatic doping \cite{huard08,khomyakov09,giovannetti08}.
In the case of SiC(0001), a similar
doping occurs from  the carbon buffer layer,
which has a high density of weakly
dispersive interface states \cite{varchon07,mattausch07}.
This can be described by a capacitor model including
quantum capacitance \cite{kopylov10},
\begin{eqnarray}
n(x)=\gamma\left(W_{sg}-\frac{e^{2}}{\epsilon_{0}}h(x)n(x)-\hbar v_{f}\sqrt{\pi n(x)}\right)
\label{chargecon}
\end{eqnarray}
where $n(x)$ is the electron density in graphene,
$W_{sg}$ is the workfunction difference between the carbon buffer layer
and graphene, $h$ is the distance between them, and
$\gamma$ is the buffer-layer density of states.
We denote this as ``Model 1''.
% JT tried moving the following to a later position:
In view of the flat bands \cite{varchon07,mattausch07},
we take the limit of large $\gamma$ and we adjust $W_{sg}$
to reproduced the known doping at $h=h_{eq}$,
corresponding to heavy n-doping \cite{bostwick07,mattausch07},
with a Fermi level $E_{f}=0.45\,$eV.
The vertical displacement $h(x)$ changes the capacitive coupling,
leading to doping variations.
Substituting the relaxed geometry $h(x)$
into Eq.\,\ref{chargecon}, we calculate these variations.
The associated potential shifts $V_{d}(x)$ are shown in Fig.\,\ref{fig4}b
for different step heights.
Increasing step height leads to larger detachment and
doping variations. This translates to an increased
${\cal R}$ as shown in Fig.\,\ref{fig4}a,
still somewhat smaller than reported experimentally,
but far larger than the scattering mechanisms previously discussed.

In studies of metal induced doping of graphene,
Khomyakov and co-workers \cite{khomyakov09,giovannetti08} reported
that Eq.\,\ref{chargecon} could not properly
describe the ab initio calculations of graphene on metals,
presumably due to quantum mechanical effects such as wavefunction
overlap and correlations
They suggested that the accuracy  of Eq.\,\ref{chargecon} can be
improved by the modification \cite{khomyakov09,giovannetti08}
\begin{eqnarray}
\nonumber
h(x) &\rightarrow& h(x)+h^{\ast}\\
W_{sg} &\rightarrow& W_{sg}+\Delta_{c}(h)
\label{dft}
\end{eqnarray}
% txns.
with $h^{\ast}$ and $\Delta_{c}(h)$ approximated as independent of
the metal species.
Since the corresponding values for SiC are not known,
and the buffer layer has a large density of states the Fermi level,
we simply use the values reported for metals in Refs. \cite{khomyakov09,giovannetti08}.

% paragraph break

This ``Model 2'' gives a stronger doping variation than the
classical electrostatic model, as show in Fig.\,\ref{fig4}c.
The corresponding resistance is also increased for Model 2,
as shown in Fig.\,\ref{fig4}a, giving excellent agreement with experiment.
Indeed, we consider the degree of quantitative agreement
between ``Model 2'' and the experimental data to be somewhat
fortuitous. Nevertheless, it is striking that, using the best
approximations available, the modulation of local doping by the
step can account for the observed resistance, while other
mechanisms are all far too small. 

% should we have a paragraph break here too?

\textcolor{black}{
In principle there could be additional electronic states associated
with the step that would change the resistance; but it is not necessary
to assume such states in order to explain the resistance.
Here we assumed the vertical surface
of the atomic step to be electrically inert.
This is reasonable since the SiC(000$\bar{1}$) surface
is electrically inert, and the extra states are associated
with the buffer layer.
In addition, our results are relatively robust against the
uncertainties in the bending stiffness and the van der Waals binding energy.
For example, if we assume an unreasonably large $E_B = 80\,$meV instead of  $E_B = 40\,$meV,
the detachment length for the largest step height decreases from $1.56\,$nm to $1.41\,$nm,
suggesting a decrease in resistance of only $\approx 10\%$.  A factor of 2 change in the assumed bending stiffness would have a similar effect.}

As seen in Fig.\,\ref{fig4}b, the graphene is almost fully 
depleted of carriers in the detached region. This suggests
a simple model where the graphene is completely undoped
over the detachment length $\ell_d$.
Our situation then resembles the problem of minimum conductivity,
often discussed in the literature\cite{twor06}.
Transport in this regime is mediated by evanescent modes,
but instead of an exponential decay the graphene bandstructure leads
to a unique linear (``pseudo-diffusive'') behavior \cite{twor06},
where ${\cal R} \approx\tfrac{ \pi^2\hbar}{2e^2}\ell_d$.
This represents the upper bound for resistance due to doping variations,
where the doping goes to zero in the detached region,
and overestimates the calculations of ``Model 2'' by about $50\%$.
Combining this with our previous result that $\ell_d \approx 1.2 h_{s}$
explains the linear scaling of resistance with step height
(i.e. ${\cal R}_{step}\propto h_{s}$) observed in experiment.

\emph{Conclusions:}
We examined the structural deformation of graphene
crossing over a substrate step, and the various intrinsic mechanisms
that may cause electron scattering at the step.
We found that deformation gives only a small electron scattering.
For graphene on SiC, where the substrate induces considerable doping,
the dominant mechanism is rather the abrupt variation in
potential and doping due to detachment of the graphene from the substrate
as it passes over a step.
Our result reconcile with the various experimental
observations, i.e. that ${\cal R}_{step}$ increases
with step density\cite{dimitrakopoulos11},
step heights\cite{shuaihua}, step bunching\cite{farmer11,bryan11},
all of them are manifestations of increasing $\ell_{d}$.

\emph{Acknowledgements}
We thank the authors of Ref.~\cite{shuaihua}
for providing experimental data prior to publication.
TL acknowledges use of a computing cluster
provided by Network for Computational Nanotechnology,
and partial funding from INDEX-NRI.

\end{document}